\documentclass[twocolumn, prl, a4paper, 10pt, superscriptaddress, nopacs, floatfix, longbibliography]{revtex4-1}
\usepackage{mathtools}
\usepackage{siunitx}
\usepackage{graphicx}
\usepackage{xcolor}
\usepackage{physics}
\usepackage{nicefrac}
\usepackage{bm}
\usepackage{float}
\usepackage[utf8]{inputenc}
\usepackage[colorlinks, citecolor=blue, linkcolor=black]{hyperref}
\usepackage{ulem}
\usepackage{booktabs}
\usepackage{multirow}
\usepackage{xspace}

\newcommand{\VB}{V$_\text{B}^-$\xspace}

\begin{document}

\title{Spin defects in hexagonal boron nitride as two-dimensional strain sensors}

\author{Z. Mu}
\affiliation{MOE Key Laboratory of Micro and Nano Systems for Aerospace, School of Mechanical Engineering, Northwestern Polytechnical University, Xi’an, China}
\affiliation{Laboratoire Charles Coulomb, Universit\'e de Montpellier and CNRS, 34095 Montpellier, France}
\author{Z. Zhang}
\affiliation{School of Sciences, Great Bay University, Dongguan 523000, China}
\affiliation{State Key Laboratory of Silicon and Advanced Semiconductor Materials, Zhejiang University, Hangzhou 310027, China}
\author{J. Frauni\'e}
\author{C. Robert}
\affiliation{Universit\'e de Toulouse, INSA-CNRS-UPS, LPCNO, 135 Avenue Rangueil, 31077 Toulouse, France}
\author{G. Seine}
\affiliation{CEMES-CNRS and Universit\'e de Toulouse, 29 rue J. Marvig, 31055 Toulouse, France}
\author{B. Gil}
\affiliation{Laboratoire Charles Coulomb, Universit\'e de Montpellier and CNRS, 34095 Montpellier, France}
\author{G. Cassabois}
\affiliation{Laboratoire Charles Coulomb, Universit\'e de Montpellier and CNRS, 34095 Montpellier, France}
\affiliation{Institut Universitaire de France, 75231 Paris, France}
\author{V. Jacques}
\email{vincent.jacques@umontpellier.fr}
\affiliation{Laboratoire Charles Coulomb, Universit\'e de Montpellier and CNRS, 34095 Montpellier, France}

\begin{abstract}
Lattice deformation is a powerful way to engineer the properties of two-dimensional (2D) materials, making their precise measurement an important challenge for both fundamental science and technological applications. Here, we demonstrate that boron-vacancy (\VB) color centers in hexagonal boron nitride (hBN) enable quantitative strain sensing with sub-micrometer spatial resolution. Using this approach, we precisely quantify the strain-induced shift of the E$_{\rm 2g}$ Raman mode in a multilayer hBN flake under uniaxial stress, establishing \VB centers as a new tool for strain metrology in van der Waals heterostructures. Beyond strain sensing, our work also highlights the unique multimodal sensing functionalities offered by \VB centers, which will be valuable for future studies of strain-engineered 2D materials.
\end{abstract} 

\maketitle
Strain engineering is a highly effective means of tuning the electronic, optical, and magnetic properties of two-dimensional (2D) materials and their associated van der Waals heterostructures~\cite{reviewStrain2,reviewStrain1,reviewStrain3}. Over the past decade, a broad variety of methodologies have been developed to achieve precise control over lattice deformation in 2D materials, including deposition onto mechanically deformable substrates, nanoindentation, or bubble formation during crystal growth~\cite{reviewStrain2}. However, fully exploiting strain engineering to tailor novel functionalities in 2D materials also requires parallel advances in techniques capable of providing independent, {\it in situ} measurements of lattice deformation with high spatial resolution. This need arises because the actual strain distribution in a 2D material often deviates from theoretical predictions owing to several effects that cannot be captured by conventional finite element simulations, such as layer slippage, substrate adhesion, edge reconstruction or interfacial and interlayer strain transmission~\cite{Mechanics2Dreview,Song2025}.

\indent In this Letter, we explore a novel approach to strain sensing in 2D materials that relies on a quantum sensing platform based on boron-vacancy (\VB) color centers in hexagonal boron nitride (hBN)~\cite{doi:10.1080/23746149.2023.2206049}. This photoluminescent point defect features a spin-triplet ground state, whose electron spin resonance (ESR) frequencies can be interrogated via optically detected magnetic resonance (ODMR) spectroscopy under ambient conditions~\cite{Gottscholl2020}, even in the limit of ultra-thin hBN layers~\cite{DurandPRL2023}. In addition, the ESR frequencies are highly sensitive to external perturbations such as magnetic and electric fields, pressure, temperature, or strain~\cite{ACSPhot_Guo2021,GottschollNatCom2021,GaoStrain2022,Udvarhelyi2023,IgorStrain2022}. These properties make  \VB centers in hBN a promising candidate for the development of flexible, multifunctional sensors that can be readily integrated into van der Waals heterostructures for probing the physics of 2D materials with atomic-scale proximity~\cite{Du2021,Tetienne2023,PhysRevApplied.18.L061002,Mu2025,HarvardArxiv}. Here, we present the first experimental study of \VB centers under continuously tunable in-plane strain, exploiting the flexibility of thin hBN flakes as a host. Using an ensemble of \VB centers, we demonstrate quantitative strain sensing  with sub-micrometer spatial resolution. We then use this capability to precisely calibrate the strain induced in a multilayer hBN flake under uniaxial stress, and we investigate the strain-induced shift of the E$_{\rm 2g}$ Raman mode. Despite the central role of hBN in 2D materials research, the impact of strain on its vibrational properties has received surprisingly little attention, with only a few studies published in the literature~\cite{Cai2017,Androulakis2018,Blundo2022}. Our results provide a precise quantification of the strain dependence of the E$_{2g}$ Raman mode in multilayer hBN, revealing a shift of $-24.7\pm2.4$~cm$^{-1}/\%$, which can serve as a reference for future studies of strain in van der Waals heterostructures. Beyond this result, our work highlights the potential of \VB centers as a robust tool for multimodal sensing in van der Waals heterostructures, with important implications for the study of strain-engineered 2D materials.

\begin{figure}[t]
  \centering
  \includegraphics[width = 8.5cm]{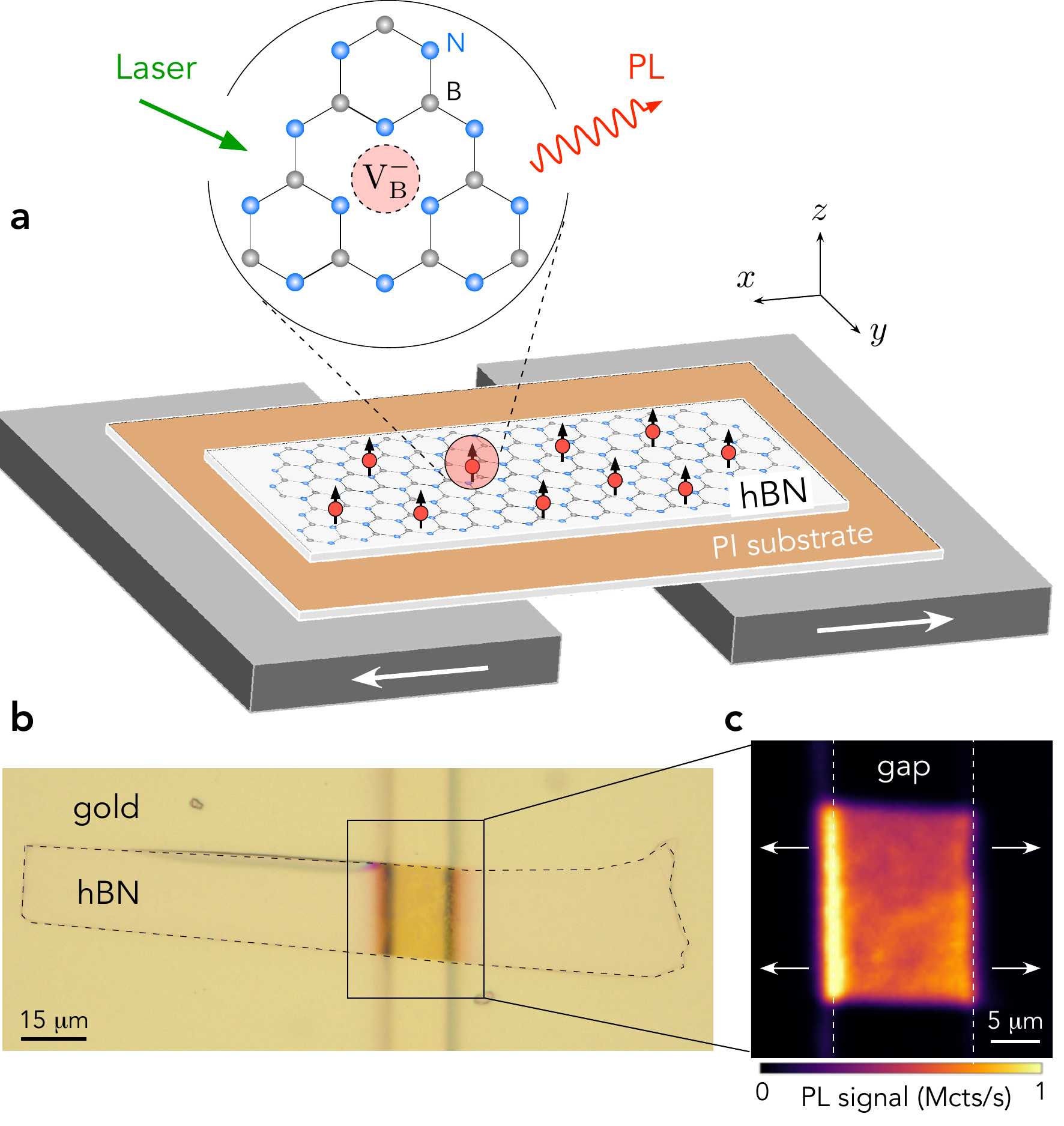}
  \caption{(a) Principle of the experiment. \VB spin defects in a thin hBN flake deposited on a strain-transmitting substrate are incorporated into a strain cell, and employed as local strain sensors. The 100-nm thick Au coating is omitted for clarity. (b) Optical image of the sample placed into the strain cell. (c) PL raster scan recorded around the gap of the sample, revealing the PL signal produced by \VB centers in hBN. The green laser excitation power is set to $1$~mW. The stretching directions are indicated by the white arrows.}
\end{figure}

\indent In our experiments, a flexible polyimide (PI) film was used as a strain-transmitting substrate. An elongated hBN flake with uniform thickness ($\sim 20$~nm) was mechanically exfoliated from a commercial hBN crystal (HQ graphene) and transferred onto the PI substrate (see~SI). A 100-nm thick Au coating was then deposited to clamp the hBN flake more firmly onto the strain-transmitting substrate, leaving only a $\sim15$~$\mu$m gap without top coating [see Fig.~1(b)]. The full stack was subsequently implanted with nitrogen at $30$~keV energy with a dose of $10^{14}$~ions/cm$^2$ to create \VB centers in the hBN flake, and finally integrated into a custom-built strain cell composed of three piezoelectric actuators mounted in parallel (see~SI)~\cite{RSI_strainCell}. Applying a positive voltage to the two outer piezoelectric actuators of the strain cell induces a stretching of the PI substrate, which effectively transfers tensile strain to the hBN flake, as schematically illustrated in Fig.~1(a).

\indent The impact of strain on the optical and spin properties of \VB centers in hBN was investigated with a scanning confocal microscope operating under ambient conditions. A laser excitation at $532$~nm was focused onto the hBN flake integrated in the strain cell through a long-working-distance objective with a numerical aperture of $0.42$. The PL signal was collected by the same objective, and guided either to a spectrometer or to a silicon avalanche photodiode operating in photon counting regime. A typical PL raster scan of a hBN flake is shown in Fig.~1(c). The PL signal features a broad spectral line in the near infrared that corresponds to the characteristic emission of \VB centers (see~SI). ODMR spectroscopy was performed by applying a microwave excitation via an external loop antenna placed in close proximity to the sample. The same setup was also used to perform Raman spectroscopy of the hBN flakes. 

We start by introducing the Hamiltonian describing the spin triplet ground state ($S=1$) of the \VB center in order to determine how its ESR frequencies evolve with strain. Neglecting the hyperfine interaction with neighboring nuclear spins, the spin Hamiltonian of the \VB center's ground state reads
\begin{equation}
\hat{\mathcal{H}}=D_0\hat{S}_z^2 + E(\hat{S}_{x}^{2}-\hat{S}_{y}^{2}) + \hat{H}_{\rm strain} \ ,
\end{equation}
where $\{\hat{S}_x,\hat{S}_y,\hat{S}_z\}$ are the dimensionless electron spin operators. The first term results from the spin-spin interaction between the two unpaired electrons of the defect, which leads to an axial zero-field splitting $D_0\sim3.47$~GHz between the $|m_s=0\rangle$ and $|m_s=\pm1\rangle$ spin sublevels~\cite{Gottscholl2020}, where $m_s$ denotes the electron spin projection along the $c$ axis~($z$) of the hBN crystal. The second term arises from the interaction of the \VB center with a local electric field produced by surrounding charges in the hBN flake~\cite{Gong2023NatCom,DurandPRL2023,Udvarhelyi2023}, which mixes the $|m_s=\pm1\rangle$ spin sublevels, yielding new eigenstates $| \pm \rangle$ separated by an orthorhombic splitting $2E$ [Fig.~2(a)]. This splitting increases with the density of charges in the hBN crystal and thus with the density of \VB centers. For the hBN flakes used in this work, $E\sim 60$~MHz, corresponding to a density of \VB centers around $100-150$ ppm~\cite{Gong2023NatCom,Udvarhelyi2023}. The last term finally describes the spin-strain interaction, which is expressed as~\cite{Udvarhelyi2023}
\begin{align}\label{strain}
&\nonumber\hat{H}_\text{strain}=
\left(\varepsilon_{xx}+\varepsilon_{yy}\right)g_{1} \hat{S}_{z}^2 
\\\nonumber&+ \left(\varepsilon_{xx}-\varepsilon_{yy}\right) \left[g_{2}(\hat{S}_{x}^{2}-\hat{S}_{y}^{2}) +g^{\prime}_{2}(\hat{S}_{x}\hat{S}_{y}+\hat{S}_{y}\hat{S}_{x})\right]
\\&+ \left(\varepsilon_{xy}+\varepsilon_{yx}\right) \left[g_{3}(\hat{S}_{x}^{2}-\hat{S}_{y}^{2}) +g^{\prime}_{3}(\hat{S}_{x}\hat{S}_{y}+\hat{S}_{y}\hat{S}_{x})\right] \ ,
\end{align}
where $g_i$ are the strain coupling coefficients and $\varepsilon_{ij}$ the components of the strain tensor. Note that the out-of-plane strain components $\{\varepsilon_{xz},\varepsilon_{yz},\varepsilon_{zz} \}$ are not included because they have a negligible impact on the in-plane hBN structure that contains the electron spin density of the \VB center. The last two terms of the spin-strain Hamiltonian yield an additional orthorhombic splitting. However, ab initio calculations have shown that the corresponding strain coupling coefficients $\{g_{2},g^{\prime}_{2},g_{3},g^{\prime}_{3}\}$ are very weak, such that the orthorhombic splitting is always dominated by the interaction with the local electric field in the strain range considered in this work ($<1 \%$)~\cite{Udvarhelyi2023}. The spin Hamiltonian can thus be simplified as
\begin{equation}
\hat{\mathcal{H}}=\left[D_0+g_{1}(\varepsilon_{xx}+\varepsilon_{yy})\right] \hat{S}_z^2 + E(\hat{S}_{x}^{2}-\hat{S}_{y}^{2})  \ ,
\end{equation}
and the ESR frequencies of the \VB center are then given by $\nu_{\pm}=D_\varepsilon\pm E$, where $D_\varepsilon=D_0+g_{1}(\varepsilon_{xx}+\varepsilon_{yy})$.
\begin{figure}[t]
  \centering
  \includegraphics[width = 8.7cm]{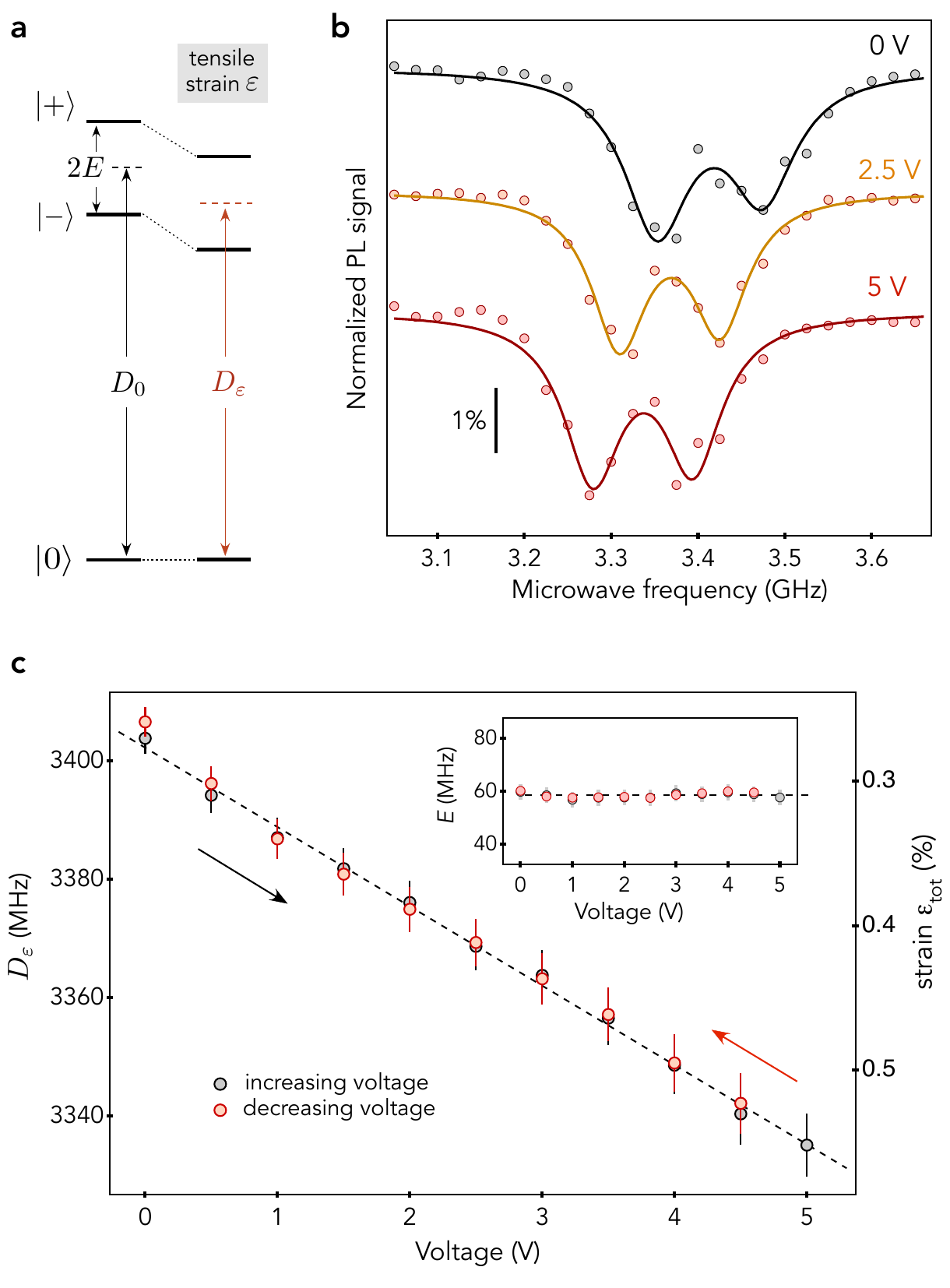}
  \caption{(a) Energy level structure of the \VB center's spin triplet ground state illustrating the impact of tensile strain. (b) ODMR spectra recorded at the center of the hBN flake [see Fig. 1(c)] for different voltages applied to the strain cell. The solid lines are data fitting with Lorentzian functions from which the parameters $D_\varepsilon$ and $E$ are extracted. (c) Strain $\varepsilon_{\rm tot}$ (right axis) transferred to the hBN flake when the voltage applied to the strain cell gradually increases from $0$ to $5$~V (black markers) and then returns back to $0$~V (red markers). Strain values are obtained by converting measurements of the zero-field splitting parameter $D_\varepsilon$ (left axis) into strain using $g_{1}=-245$~MHz/\% and $D_0=3.47$~GHz. The error bars correspond to the uncertainty on the strain values, which mainly arise from the uncertainty ($\sim 4\%$) on the coupling parameter $g_1$. The dashed lines is data fitting with a linear function. The inset shows the evolution of the orthorhombic splitting $E$ with voltage.}
\end{figure}

First-principles calculations predict a strain coupling coefficient of $g_{1,\rm ML}=-192$~MHz/\% for a hBN monolayer, which increases to $g_{1}=-245\pm10$~MHz/\% for \VB centers embedded in multilayer hBN flakes such as those used in this work~\cite{Udvarhelyi2023}.
Tensile strain (taken positive by convention) thus leads to a reduction of the {\it effective} zero-field splitting parameter $D_\varepsilon$~[Fig.~2(a)]. These theoretical predictions have been quantitatively confirmed by several recent experiments performed on multilayer hBN flakes under hydrostatic stress conditions~\cite{Mu2025,HarvardArxiv,ChinaArxiv}. We note that the strain coupling coefficient $g_{1}$ is one order of magnitude smaller than that reported in earlier works~\cite{GottschollNatCom2021,GaoStrain2022}. In those studies, the strain susceptibility of the \VB center was indirectly inferred from the temperature dependence of the axial zero-field splitting parameter, assuming it arises solely from an effective stress due to thermal distortion of the hBN lattice. More recent work has shown that the variations of the zero-field splitting parameter with temperature instead originate from temperature-dependent spin-phonon interactions~\cite{PhysRevB.111.024108}, explaining the discrepancy in the reported strain susceptibilities. In the following, we employ the theoretically predicted and experimentally confirmed strain coupling coefficient $g_{1}$ to extract the local strain $\varepsilon_{\rm tot}=(\varepsilon_{xx}+\varepsilon_{yy})$ in a multilayer hBN flake from measurements of the ESR frequencies.

\indent ODMR spectra recorded at the center of the hBN flake for different voltages applied to the strain cell are shown in Fig.~2(b). At zero voltage, we obtain $D_\varepsilon\sim3.40$~GHz, corresponding to an initial tensile strain $ \varepsilon_{\rm tot}\sim 0.3\%$, which is likely transferred to the hBN flake when it is fixed on the strain cell. Raising the applied voltage results in an increased strain, which gradually shifts the magnetic resonances of the \VB center to lower frequencies. For voltages up to $5$~V, the strain transferred to the hBN flakes increases linearly [Fig.~2(c)] and the process is perfectly reversible, the initial strain being recovered when the voltage is swept back to $0$~V. In this voltage range, the hBN flake thus follows an elastic deformation behavior.

\VB spin sensors can also be used to infer the strain distribution within the hBN flake with a diffraction-limited spatial resolution ($< 1 \ \mu$m). Figure 3 shows typical strain profiles recorded along the long axis ($x$) of the hBN flake for various voltages applied to the strain cell. These measurements reveal a slight strain gradient of about $5\times10^{-3}$~\%/$\mu$m, which emerges between the two sides of the strain cell. This gradient likely results from different adhesion of the hBN flake on the strain-transmitting substrate at the opposite edges of the gap.
\begin{figure}[b]
  \centering
  \includegraphics[width = 8.6cm]{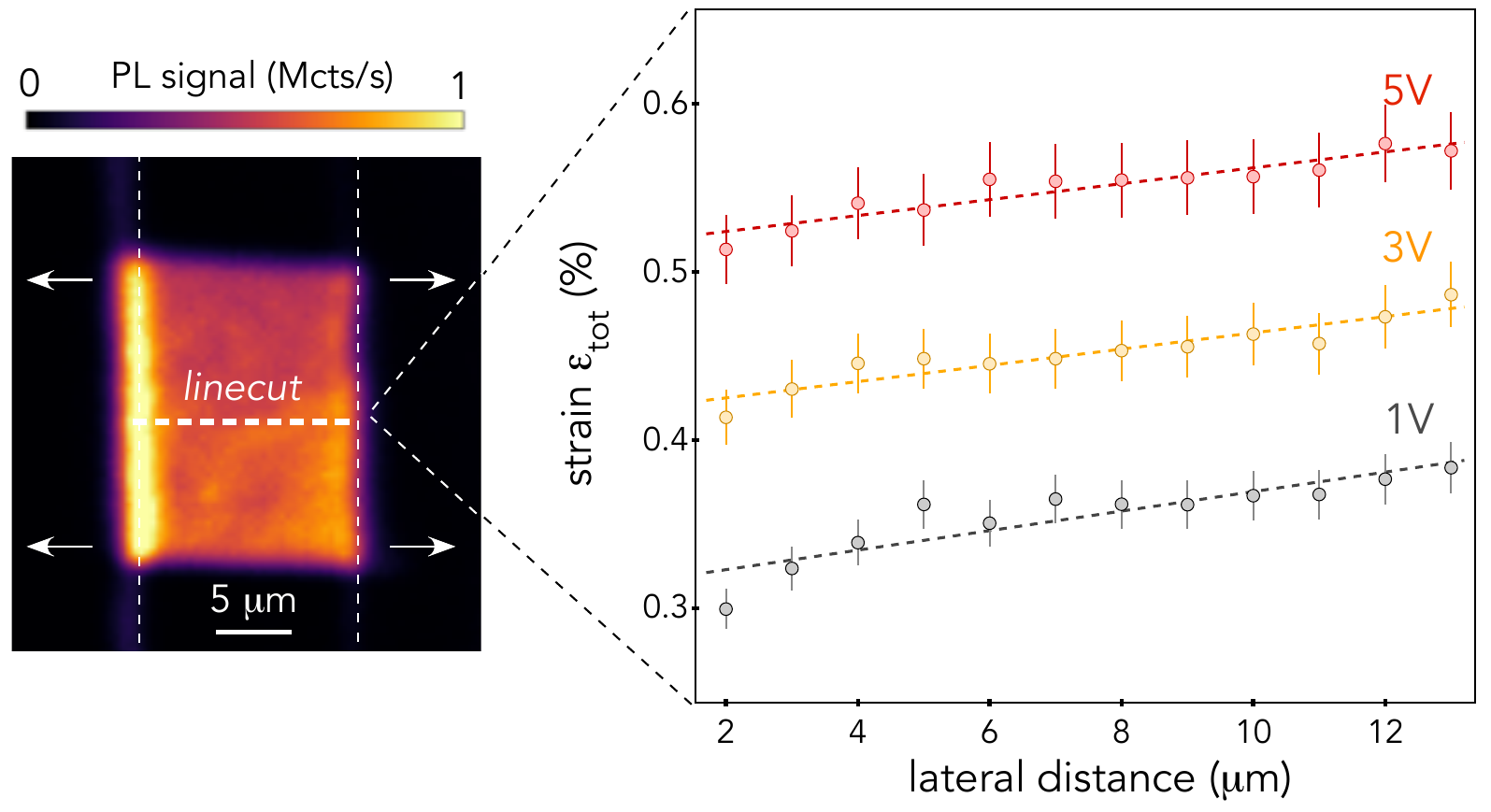}
  \caption{Strain ($\varepsilon_{\rm tot}$) profiles measured along the linecut shown in the PL scan (left panel) for different voltages applied to the strain cell. The dashed lines are data fitting with a linear function from which we extract a strain gradient of about $\sim5\times10^{-3}$~\%/$\mu$m for the three strain profiles. }
\end{figure}

Several remarks have to be made at this stage. First, the orthorhombic splitting is not modified in the considered strain range [see inset in Fig.~2(c)], supporting our assumption that this parameter is dominated by the interaction with a local electric field. Second, the PL signal of \VB centers and the contrast of ODMR spectra are not significantly modified when tensile strain increases [see Fig.~2(b)]. These findings might seem in contradiction with recent studies on the optical and spin properties of \VB centers placed under hydrostatic pressure conditions in a diamond anvil cell (DAC), which reported a drastic reduction of both PL signal and ODMR contrast for pressures above a few GPa~\cite{Mu2025,HarvardArxiv,ChinaArxiv}. A key difference between experiments carried out within a  DAC and with a strain cell is the presence or not of an out-of-plane ($z$) stress component. Although this component has a negligible impact on the ground state electron spin sublevels of the \VB centre~\cite{Udvarhelyi2023}, it likely modifies the structure of its excited states and the spin-dependent dynamics of optical cycles~\cite{ivady2020,Clua2024}. Experiments performed under hydrostatic pressure in a DAC suggest that compressive out-of-plane stress results in a quenching of both PL signal and ODMR contrast owing to variations of non-radiative decay rates via a metastable singlet state during optical cycles~\cite{Mu2025}. These effects are likely not relevant to the present work because stress is applied in-plane. It would be interesting to study in future the regime of tensile, out-of-plane stress, which could lead to an enhancement of the PL signal, possibly providing new insight into the origin of the intrinsic dimness of the \VB center emission. 

As a first application of \VB-based strain sensing, we now investigate the strain-induced variation of the E$_{\rm 2g}$ Raman mode in hBN, a topic that remains scarcely documented in the literature despite the central role of hBN in 2D materials research. To this end, both Raman and ODMR spectra are sequentially measured at the center of the hBN flake while gradually increasing the voltage applied to the strain cell. The ODMR spectra provide a precise, {\it in~situ} strain calibration at the exact location where Raman spectra are recorded. As shown in Fig.~4(a), the effective zero-field splitting parameter of the \VB center initially decreases linearly with the applied voltage, consistently with the previous experiment. It reaches a minimum value $D_{\varepsilon}\sim 3.32$~GHz at a voltage of $\sim 7$~V, corresponding to a maximum strain of $\varepsilon_{\rm tot}\sim 0.6\%$. Interestingly, further elongation of the strain cell does not increase the strain in the hBN flake, but instead leads to its gradual release as the voltage rises from $7$~V to $10.5$~V. This effect most likely result from slippage between the multilayer hBN flake and the PI substrate~\cite{Androulakis2018}. We note that the ODMR linewidth is not modified in this regime, suggesting that in-plane strain does not vary significantlly across the thickness of the hBN flake. When the voltage is swept back to $0$~V, the strain decreases linearly with a slope similar to that observed during the voltage ramp-up and reaches a compressive state $\varepsilon_{\rm tot}\sim-0.1\%$ at zero voltage [Fig.~4(a)]. This pronounced hysteretic behavior of strain in a hBN flake under uniaxial stress highlights the complexity of the mechanical response in layered 2D materials, which cannot be captured by standard numerical simulations.
\begin{figure}[t]
  \centering
  \includegraphics[width = 8.7cm]{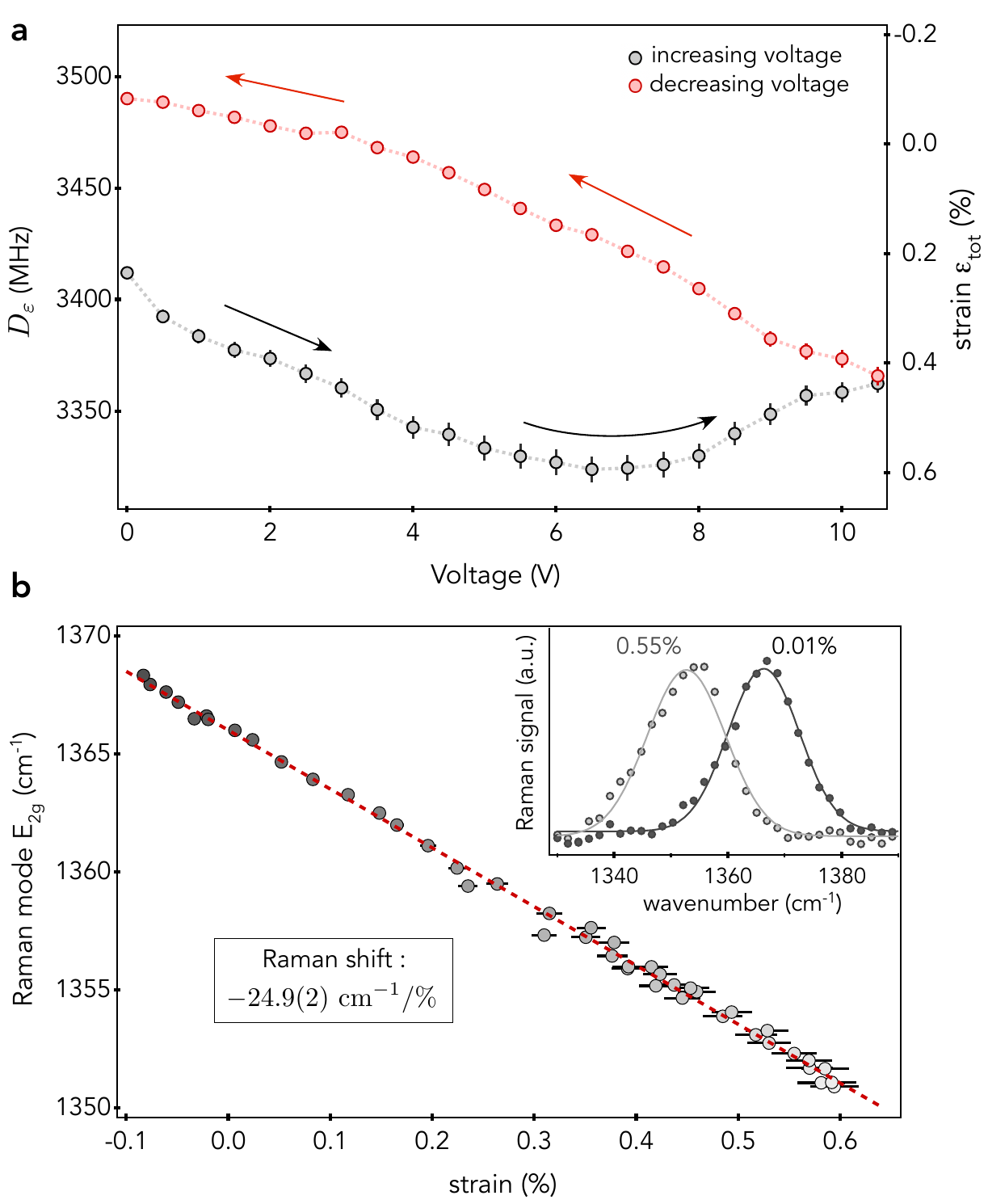}
  \caption{(a) Evolution of the strain $\varepsilon_{\rm tot}$ (right axis) in the hBN flake when the voltage applied to the strain cell gradually increases from $0$ to $10.5$~V (black markers) and then returns back to $0$~V (red markers). These values are obtained by converting measurements of the zero-field splitting parameter $D_\varepsilon$ (left axis) into strain using $g_{1}=-245$~MHz/\% and $D_0=3.47$~GHz. The error bars correspond to the uncertainty on the strain values, which mainly arise from the uncertainty ($\sim 4\%$) on the coupling parameter $g_1$. (b) Frequency of the E$_{\rm 2g}$ Raman mode in hBN as a function of in-plane strain $\varepsilon_{\rm tot}$. Data fitting with a linear function (red dashed line) yields a strain-induced shift of the E$_{\rm 2g}$ Raman-mode of $-24.9(2)$~cm$^{-1}/\%$. The inset shows typical Raman spectra recorded for two different values of $\varepsilon_{\rm tot}$. The width of the Raman signal ($\sim 15$~cm$^{-1}$) is not significantly modified with applied stress.}
\end{figure}

\indent With the strain precisely calibrated using \VB centers, we proceed with the analysis of the Raman spectra. At zero strain, the E$_{\rm 2g}$ Raman mode of hBN is observed at $\omega_0\sim1366$~cm$^{-1}$ and shifts linearly with strain, with a slope $\Delta_{\rm 2g}=-24.9(2)$~cm$^{-1}/\%$ [Fig.~4(b)], corresponding to a Gr$\ddot{\rm u}$neisen parameter $\gamma_{\rm 2g}=1.82(5)$. In addition to the overall shift of the Raman mode, the anisotropic strain distribution in the hBN flake is also expected to split the doubly-degenerate E$_{\rm 2g}$ mode owing to a lowered crystal symmetry~\cite{Androulakis2018}. This splitting cannot be observed in our experiments, most likely because of the broad Raman linewidth ($\sim 15$~cm$^{-1}$) resulting from the implantation of the hBN flake with nitrogen atoms [see inset in Fig.~4(b)]. From similar measurements performed on other hBN flakes subjected to uniaxial stress, we infer an average shift of the E$_{\rm 2g}$ Raman mode $\bar{\Delta}_{\rm 2g}= -24.7 \pm 2.4$~cm$^{-1}/\%$ and an average Gr$\ddot{\rm u}$neisen parameter $\bar{\gamma}_{\rm 2g}=1.80\pm0.17$ (see SI). These values, which are fully consistent with the few previous reports available in the literature~\cite{Cai2017,Androulakis2018,Blundo2022}, will likely serve as a reference for future studies of strain in van der Waals heterostructures.

Beyond strain sensing, \VB centers in hBN enable simultaneous measurements of magnetic fields, a feature that could prove useful for exploring strain-controlled magnetic properties of 2D magnets~\cite{Miao2020,Cenker2022}. The impact of strain and magnetic field on the ESR frequencies can be easily disentangled by applying a bias magnetic field $B_b$ along the $c$ axis ($z$) of hBN. In the limit $B_b\gg E/\gamma_e$, the ESR frequencies are expressed as $\nu_{\pm}=D_\varepsilon\pm\gamma_eB_b$, where $\gamma_e=28$~GHz/T is the electron spin gyromagnetic ratio. From an ODMR spectrum, strain can thus be inferred from the sum of the two ESR frequencies, while their difference directly yields the magnetic field variation. The shot-noise limited magnetic field sensitivity is given by $\eta_{B}\approx 0.7 \times \frac{1}{\gamma_e}\times\frac{\Delta\nu}{\mathcal C\sqrt{\mathcal R}}$, where $\mathcal{C}$ is the contrast of the ODMR spectrum, $\Delta\nu$ its linewidth, and $\mathcal{R}$ the rate of detected photons~\cite{PhysRevB.84.195204}. Replacing $\gamma_e$ by the strain coupling coefficient $g_{1}$ yields the sensitivity $\eta_{\varepsilon}$ to strain. For the 20-nm-thick hBN flake studied in this work, we estimate a strain sensitivity $\eta_{\rm \varepsilon}\sim 0.03\%/\sqrt{\rm Hz}$ together with a magnetic field sensitivity $\eta_{\rm B}\sim 250 \ \mu$T/$\sqrt{\rm Hz}$ (see SI). These performances could be further improved by at least one order of magnitude by optimizing the microwave excitation used for ODMR spectroscopy~\cite{Debasish2025} and by relying on isotopically purified hBN crystals~\cite{Clua2023,Gong2024}.\\

To conclude, we have demonstrated that \VB centers in hBN provide quantitative {\it in situ} strain measurements with sub-micrometer spatial resolution, and we used this capability to precisely quantify the strain-induced shift of the $E_{\rm 2g}$ Raman mode. These results establish \VB centers as a new tool for strain metrology in van der Waals heterostructures. Compared to other strain sensing methods such as Raman spectroscopy, a key advantage of \VB centers in hBN lies in their ability to simultaneously probe several external perturbations with high sensitivity. This unique multimodal sensing functionality opens up exciting opportunities to investigate the impact of lattice deformations on the vibrational, magnetic, and electronic properties of 2D materials.\\

\noindent {\it Acknowledgements} - This work was supported by the French Agence Nationale de la Recherche through the project Qfoil (ANR-23-QUAC-0003), the program ESR/EquipEx+ 2DMAG (grant ANR-21-ESRE-0025), the project NanoX Q2D-SENS (grant ANR-17-EURE-0009) in the framework of the “Programme des Investissements d’Avenir”, and the Institute for Quantum Technologies in Occitanie. Z.M. acknowledges support from the National Science Foundation of China No. 52575671 and the Fundamental Research Funds for the Central Universities. Z.Z. acknowledges support from the National Natural Science Foundation of China (12404124), the Guangdong Basic and Applied Basic Research Foundation (2023A1515110419), the Pearl River Talent Recruitment Program (2023QN10X156), and the Open Project of State Key Laboratory of Silicon and Advanced Semiconductor Materials, Zhejiang University (SKL2025-12).


%

\end{document}